\documentclass[aip,amsmath,amssymb,reprint]{revtex4-1}

\usepackage{graphicx}
\usepackage{dcolumn}
\usepackage{bm}

\usepackage[utf8]{inputenc}
\usepackage[T1]{fontenc}
\usepackage{mathptmx}

\usepackage{bm}
\usepackage{bbm}
\usepackage{amsfonts}
\usepackage{amsmath}
\usepackage{natbib}
\providecommand{\keywords}[1]{\textbf{\textit{Index terms---}} #1}
\usepackage[mathlines]{lineno}
\usepackage[colorinlistoftodos]{todonotes}
\usepackage[colorlinks=true, allcolors=blue]{hyperref}

\begin{document}

\newcommand{\unamone}{Departamento de Sistemas Complejos, Instituto de Fisica,
Universidad Nacional Aut\'onoma de M\'exico, Apartado Postal 20-364,01000,
Ciudad de M\'exico, M\'exico.}
\newcommand{\unamtwo}{Instituto de F\'isica,
Universidad Nacional Aut\'onoma de M\'exico, Apartado Postal 20-364 01000,
Ciudad de  M\'{e}xico, M\'{e}xico}
\newcommand{\uama}{\'Area de F\'isica Te\'orica y Materia Condensada,
Universidad Aut\'onoma Metropolitana Azcapotzalco, Av. San Pablo 180,
Col. Reynosa-Tamaulipas, 02200 Cuidad de M\'exico, M\'exico}

\preprint{AIP/123-QED}

\title[Sample title]{Floquet spectrum for anisotropic  and tilted Dirac materials
under linearly polarized light at all field intensities}

\author{J. C. Sandoval-Santana}
\email{jcarlosss@fisica.unam.mx}
\affiliation{\unamtwo}
\author{V. G. Ibarra-Sierra}
\email{vickkun@fisica.unam.mx}
\affiliation{\unamone}
\author{A. Kunold}
\email{akb@azc.uam.mx}
\affiliation{\uama}
\author{Gerardo G. Naumis}
\email{naumis@fisica.unam.mx}
\affiliation{\unamone}

\date{\today}

\begin{abstract}
The Floquet spectrum in an anisotropic tilted Dirac semimetal 
modulated by linearly polarized light is addressed through the
solution of the time-dependent Schr\"odinger equation for the
two-dimensional Dirac Hamiltonian via the Floquet theorem.
The time-dependent wave functions and the quasienergy spectrum
of the two-dimensional Dirac Hamiltonian under the normal incidence
of linearly polarized waves are obtained for an arbitrarily intense
electromagnetic radiation. We applied a set of unitary transformations
to reduce the Schr\"odinger equation to an ordinary second-order
differential Hill equation with complex coefficients.
Through the stability analysis of this differential equation,
the weak and strong field regimes are
clearly distinguished in the quasi-spectrum. In the
weak electric field regime, above a certain threshold given by the field parameters, the spectrum mostly resembles
that of free electrons in graphene.
Below this threshold, in the strong electric field regime,
the spectrum abruptly becomes highly anisotropic and a gap opens up.
As an example,  we apply the results to the particular case of borophene.

\end{abstract}

\maketitle

\section{Introduction}

The superior physical, mechanical an chemical properties
of two-dimensional (2D) materials makes them
an ideal playground to study new and exciting kinds
of quantum phases
\cite{novoselov2012roadmap,peng2016electronic,wehling2014dirac,
ferrari2015science,Mehboudi5888Strain,Villanova2016Spin,villanova2017engineering}.
Their remarkable electronic and optical features
have intensely driven the development
of novel and innovative
optoelectronic devices \cite{bonaccorso2010graphene,bao2017graphene,ponraj2016photonics}
as broad band optical modulators
\cite{liu2011graphene,sorianello2018graphene,hao2019experimental},
solar cells
\cite{yin2014graphene,o2019graphene},
infrared photodetectors
\cite{safaei2019dirac}
and hybrid plasmonic devices
\cite{grigorenko2012graphene,fan2019graphene}.

In the last decade persistent efforts have been made
to harness the unique features of graphene's so-called
dressed electrons
\cite{lopez2008analytic,lopez2010graphene,bonaccorso2010graphene,
kibis2010metal,
Foa2011Tuning,
sun2012ultrafast,
kibis2015,
kristinsson2016control,
kibis2017all,
kibis2018electromagnetic,safaei2019dirac}
to design different kinds of optoelectronic devices.
Electromagnetic dressing, attained when
electrons strongly couple to electromagnetic fields,
substantially renormalize the energy and the velocities
in graphene\cite{kibis2018electromagnetic}.
In turn, renormalized parameters highly
depend on the light polarization: in graphene circularly polarized fields
open a dynamical gap in the Dirac point while linearly polarized fields
live it intact. Particularly, dressed electrons under linearly polarized
light induce an anisotropy of the electron dispersion relation \cite{kibis2017all}.
Electromagnetic dressing could, therefore, be used to tune the electronic
and optical
properties of graphene, including band gap and carrier
velocities\cite{kibis2018electromagnetic} which are clearly manifest in
several measurable physical properties as the photocurrent
\cite{sun2012ultrafast,Chan2017Photocurrents}.
A similar approach has been adopted in ultrafast material science
promising optical and
mechanical control of the physical properties of 
2D materials\cite{naumis2017electronic}.
In this case, it is time-dependent
strain that acts as  pseudo-electromagnetic field
\cite{carrillo2018band}.

In graphene, the weak-field regime, where light-matter coupling
is perturbative\cite{higuchi2017light,Mishchenko2018Effect,herrera2020electronic},
is well understood.
It may be pictured as a quantized-photon field interacting
with massless Dirac fermions having a conic dispersion relation.
However, in the intense-field regime, this perturbative expansion cannot
converge. The Dirac dispersion relation is highly distorted
as a result of the electronic dressing rendering the quantized-photon
picture invalid.
It is clear then that this emerging field requires new
theoretical tools that go beyond the conventional techniques
\cite{BogdanFernandez1989,Mielnik1989}.
There are numerous proposals to approach
this problem
\cite{lopez2008analytic,lopez2010graphene,kibis2010metal,
Foa2011Tuning,kibis2015,kristinsson2016control,kibis2017all},
but, even for graphene, some aspects of the interaction
between carriers and strong time-driven fields remain elusive.

Another essential question that remains to be fully answered
is how such an intense electromagnetic radiation would affect
other Dirac materials\cite{Goerbig2008Tilted,Feng2018Feng}
as borophene or black-phosphorus
\cite{mehboudi2016two,utt2015intrinsic,Poudel2019Group,Jung2020},
for example.
Particularly borophene is a remarkable anisotropic material
\cite{peng2016electronic,verma2017effect,Villanova2016Spin,Champo2019,Villanova2016Spin}.
After being theoretically
predicted thirty years ago\cite{BOUSTANI1997355},
it was until 2015 that it was synthetized  \cite{zhang2017two}.
Borophene turns out to be stronger and even more flexible than graphene.
It is a good conductor of both electricity and heat and it
is expected to be a superconductor \cite{mannix2018borophene}
with relatively high transition temperatures.

In a series of recent papers, we have investigated time-driven
anisotropic Dirac Hamiltonians as the one that describes borophene 
\cite{Zabolotskiy2016,verma2017effect,Champo2019,ibarra2019dynamical}. 
However, the mathematical complexity of the problem required several
approximations that are only valid for very intense 
fields\cite{Champo2019,ibarra2019dynamical}.
Therefore, the critical link between
weak fields, treatable with perturbation theory,
and strong fields is still missing.
In particular, the cases of linearly and elliptically polarized light were
studied in the intense field regime \cite{Champo2019,ibarra2019dynamical},
where the rather convoluted Hill equation with time dependent coefficients
was simplified into the very well known Mathieu equation with constant
coefficients.
Nevertheless, the complete study of the Hill equation entails
the interplay between the strong and weak field intensity regimes,
key to comprehend the formation of the quasienergy spectrum.
This effect can be thoroughly studied by analyzing
the quasi-energy spectrum using the Floquet theory
\cite{dittrich1998quantum,Fernandez1992,Mielnik2016,hubener2017creating,Oka2019Floquet}.

In this paper we address the general problem of a particle that
obeys the  anisotropic Dirac Hamiltonian subject to linearly
polarized light with an arbitrarily large field intensity.
The calculation of the quasienergy spectrum and the wave function
is achieved through a set of unitary transformations
that enables to reduce the matrix differential equation
into a scalar differential equation.
This, in turn, is readily solved via the Floquet theorem
and a Fourier spectral decomposition of the periodical part
of the solution.
It is shown that in the intense field regime the quasienergy
spectrum abruptly develops an anisotropic structure, absent
in the weak field regime.

The paper is organized as follows.
In Sec. \ref{sec:DiracHamiltonian} we introduce the low-energy effective 
two-dimensional anisotropic Dirac Hamiltonian and we study
the case of borophene subject to an arbitrarily intense
linearly polarized electromagnetic field.
Subsequently, in Sec. \ref{sec:QuasienergySpectrumWaveFunction},
we analyze the quasienergy spectrum that emerges from the
Hill equation by means of the Floquet approach.
Also, we find the time-dependent wave functions.
Finally, we summarize and conclude in Sec. \ref{sec:Conclusions}.

\section{Two-dimensional electrons in a tilted Dirac cone
subject to electromagnetic fields}\label{sec:DiracHamiltonian}

\subsection{The anisotropic Dirac Hamiltonian}

We start by considering a low-energy anisotropic Dirac Hamiltonian
close to one of the Dirac points.
In the particular case of $8-Pmmn$ borophene  it is given by 
\cite{Zabolotskiy2016,verma2017effect,Champo2019,HerreraNaumisKubo2019}
\begin{equation}\label{ec:AnsotropicDiracHamiltonian}
\hat{H}= \hbar v_t k_y\hat{\sigma}_0
+\hbar\left[v_x k_x\hat{\sigma}_x +v_y k_y\hat{\sigma}_y\right],
\end{equation}
where $ k_x$ and $k_y$ are the components of the two-dimensional
momentum vector $\boldsymbol{k}$,  $\hat{\sigma}_x$
and $\hat{\sigma}_y$ are the Pauli matrices,
and $\hat{\sigma}_0$ is the $2\times 2$ identity matrix.
The three velocities in the anisotropic $8-Pmmn$ borophene Dirac
Hamiltonian (\ref{ec:AnsotropicDiracHamiltonian})
are given by $v_x=0.86v_F$, $v_y=0.69v_F$ and $v_t=0.32v_F$ where
$v_F=10^6\,m/s$ \cite{Zabolotskiy2016}
is the Fermi velocity.
In Eq. (\ref{ec:AnsotropicDiracHamiltonian}), the
last two terms give rise to the familiar form of the
kinetic energy leading to the Dirac cone and
the first one tilts the
Dirac cone in the $k_y$ direction.
These two features are contained in the
energy
dispersion relation\cite{ibarra2019dynamical}
\begin{equation} \label{ec:EnergyDispersion}
E_{\eta,k}=\left(\frac{v_t}{v_y}\right) \tilde{k}_y +\nu\epsilon,
\end{equation}
where
\begin{equation}\label{ec:EpsilonCoefficient}
\epsilon=\sqrt{\tilde{k}_{x}^{2}+\tilde{k}_{y}^{2}},
\end{equation}
and $\nu= \pm 1$ is the band index.
In Eq. \eqref{ec:EnergyDispersion}, we used the set of
renormalized moments
$\tilde{k}_x=\hbar v_x k_x$, $\tilde{k}_y=\hbar v_y k_y$.
The corresponding free electron wave function is,
\begin{equation}\label{ec:WaveFunctionFree}
\psi_{\nu}(\boldsymbol{k})
  = \frac{1}{\sqrt{2}}
  \left[
  \begin{array}{c}
     1 \\
     \nu\exp(i \theta_{\boldsymbol{k}} ) 
    \end{array} \right]
\end{equation}
where $\theta_{\boldsymbol{k}}= \tan^{-1} (\tilde{k}_y/\tilde{k}_x)$.

\subsection{Linearly polarized waves} 

Now we consider a charge carrier, described by the two-dimensional
anisotropic Dirac Hamiltonian, subject to an electromagnetic wave that 
propagates along a direction perpendicular to the surface of the crystal.
The effects of the electromagnetic field
are introduced in the Hamiltonian
\eqref{ec:AnsotropicDiracHamiltonian} through the Peierls 
substitution\cite{dey2018photoinduced} 
$\hbar\boldsymbol{k}\rightarrow\hbar\boldsymbol{k}-e\boldsymbol{A}$
where $\boldsymbol{A}=(A_x,A_y)$ is the vector potential
of the electromagnetic wave.
A considerable simplification
can be achieved by adopting
a gauge in which $\boldsymbol{A}$ only
depends on time. 
The Hamiltonian \eqref{ec:AnsotropicDiracHamiltonian}
is thus transformed into
\begin{multline} \label{ec:DiracHamiltonianUnderElectromagneticField} 
\hat{H}= \frac{v_t}{v_y}\left(\tilde{k}_y-e v_y A_y\right)\hat{\sigma}_0\\
 +\left(\tilde{k}_x-e v_x A_x\right)\hat{\sigma}_x 
 +\left(\tilde{k}_y-e v_y A_y\right)\hat{\sigma}_y.
\end{multline}
Assuming that the electromagnetic wave is
linearly polarized, the vector potential
can be written as
\begin{equation}\label{ec:VectorPotential}
\boldsymbol{A}=\frac{E_0}{\Omega}\cos(\Omega t)\boldsymbol{\hat{r}},
\end{equation}
where $\boldsymbol{\hat{r}}=(1,0)$ is the polarization vector,
$E_0$ is the uniform amplitude of the electric field
and $\Omega$ is the angular frequency of the electromagnetic wave.
Observe that here the field $\boldsymbol{A}$ is not quantized and
is treated clasically. Thus, our results are valid for a field
with a large number of photons, which can be represented
by a quantum coherent field. 
In the Schr\"odinger equation corresponding to
\eqref{ec:DiracHamiltonianUnderElectromagneticField},
\begin{equation}\label{ec:DiracEquationOne}
i \hbar \frac{d}{d t} {\boldsymbol{\Psi}}(t)
  =\hat{H}{\boldsymbol{\Psi}}(t),
\end{equation}
the two dimensional spinor can be expressed as
$\boldsymbol{\Psi}(t)
 =\left(\Psi_{A}(t),\Psi_{B}(t)\right)^{\top}$, where
$A$ and $B$ label the two sublattices.

The main difficulty in
deducing the wave function's explicit form
resides in that the Hamiltonian
\eqref{ec:DiracHamiltonianUnderElectromagneticField}
couples the differential equations for
the $\Psi_{A}(t)$ and $\Psi_{B}(t)$
spinor components due to
the terms that are proportional
to $\hat\sigma_x$ and $\hat\sigma_y$.
To uncouple the spinor components we proceed
as follows.
First, applying a $45^\circ$ rotation around the $k_y$ axis
of the form
\begin{equation} \label{ec:SolutionOne}
\boldsymbol{\Psi}(t)
=\exp\left[-\frac{i}{\hbar}\left(\frac{\pi}{4} 
\right)\hat{\sigma}_y\right]\boldsymbol{\Phi}(t),
\end{equation}
conveniently transforms the non-diagonal $\hat\sigma_x$
matrix into $\hat\sigma_z$.
Indeed, substituting (\ref{ec:SolutionOne}) into 
Eq. (\ref{ec:DiracEquationOne}) yields
\begin{equation}
\label{ec:Schrokxky}
i \frac{d}{d\phi}\boldsymbol{\Phi}(\phi)
=\frac{2}{\hbar \Omega}\left[\left(\frac{v_t}{v_y}\right) 
\tilde{k}_{y} \hat{\sigma}_0+ \tilde{\Pi}_x\hat{\sigma}_z
+ \tilde{k}_{y} \hat{\sigma}_y \right]\boldsymbol{\Phi}(\phi) \, ,
\end{equation}
where the only non-diagonal remaining term
is the one proportional to $\hat\sigma_y$.
In the foregoing equation, $\phi= \Omega t/2$,
$\tilde{\Pi}_x=\tilde{k}_x-\zeta_x \cos(2 \phi)$ and
$\zeta_x=ev_xE_x/\Omega$.
The spinor components of
$\boldsymbol{\Phi}(\phi)=\left(\Phi_{+}(\phi),\Phi_{-}(\phi)\right)^{\top}$
are given by
$\Phi_{+}(\phi)=[\Psi_A(\phi)+\Psi_B(\phi)]/\sqrt{2}$ and 
$\Phi_{-}(\phi)=[\Psi_A(\phi)-\Psi_B(\phi)]/\sqrt{2}$.
Second, the term proportional to $\hat{\sigma}_0$
in  Eq. \eqref{ec:Schrokxky}
is removed by adding a time-dependent
phase to the wave function
\begin{equation}\label{ec:suprimdiag}
\boldsymbol{\Phi}(\phi)=\exp\left[-2i\left(\frac{v_t}{v_y}\right)\frac{ \tilde{k}_y}{\hbar \Omega}\, \phi\, \hat{\sigma}_0\right]\bm{\chi}(\phi),
\end{equation}
where $\bm{\chi}(\phi)=(\chi_{+1}(\phi),\chi_{-1}(\phi))^{\top}$. 
Finally, after inserting Eq. (\ref{ec:suprimdiag}) into Eq.
(\ref{ec:Schrokxky}), we follow the
procedure shown in Appendix \ref{AppendixA}.
The resulting diferential equation
takes on the form
of the Hill equation \cite{magnus2013hill}
\begin{equation}
\bm{\chi}''(\phi)+\mathbbm{F}(\phi)\bm{\chi}(\phi)=0 \, , 
\label{ec:Hilluncoupled}
\end{equation}
where the matrix $\mathbbm{F}(\phi)$ is defined as
\begin{multline}\label{ec:HillFinal}
\mathbbm{F}(\phi)=\left[a+q_1\cos(2\phi)
+q_2\cos(4\phi)\right]\hat{\sigma}_{0}\\+q_3\sin(2\phi)\hat{\sigma}_{z}.  
\end{multline}
The Hill equation parameters
\begin{eqnarray}
a&=&\left(\frac{2}{\hbar\Omega}\right)^{2}
\left(\epsilon^2+\frac{\zeta^2_{x}}{2}\right)\label{ec:Hill_a},\\
q_1&=&-8\left(\frac{\tilde{k}_x}
{\hbar\Omega}\right)\left(\frac{\zeta_{x}}{\hbar\Omega}\right),\label{ec:Hill_q1}\\
q_2&=&2\left(\frac{ \zeta_{x}}{\hbar\Omega}\right)^{2},\label{ec:Hill_q2}
\\
q_3&=& 4 i\left(\frac{\zeta_x}{\hbar\Omega}\right).\label{ec:Hill_q3}
\end{eqnarray}
are expressed in terms of the ratios of the characteristic energies
of the system.
Thereby, $\epsilon/\hbar\Omega$
is the ratio of the electron kinetic energy to the photon energy,
$\zeta_{x}/\hbar\Omega$  is the ratio of
the work done on the charged carries by the electromagnetic wave
to the photon energy and $\tilde{k}_x/\hbar\Omega$
is the ratio of the $x$ part of the electron kinetic energy
to the photon energy.

Expressing \eqref{ec:Hilluncoupled} as
a second order differential equation
is quite advantageous for the calculations that follow.
First, the evolution operator that propagates 
the state $\chi$ in time must be diagonal
since $\mathbbm{F}(\phi)$ is solely composed of the diagonal
matrices $\hat\sigma_0$ and $\hat\sigma_z$.
As a consequence of this,
the scalar differential
equations for the $\chi_{+1}(\phi)$ and $\chi_{-1}(\phi)$
spinor components decouple. Moreover, the differential
equation for the $\chi_{-1}(\phi)$ component turns out to be
the complex conjugate of the one for $\chi_{+1}(\phi)$.
Both differential equations may be summarized by
\begin{multline}\label{ec:HillCompoentEta}
\chi_{\eta}''(\phi)+\left[a+ q_1\cos(2\phi)+q_2\cos(4\phi)\right.\\
+\left. \eta q_3 \sin(2\phi)\right]\chi_{\eta}(\phi)=0,    
\end{multline}
where $\eta=\pm 1$.

This alternative form of the Schr\"odinger equation considerably
simplifies the computation and analysis of the stability spectrum. 

\begin{figure}[t]
\centering
\includegraphics[width=.43\textwidth]{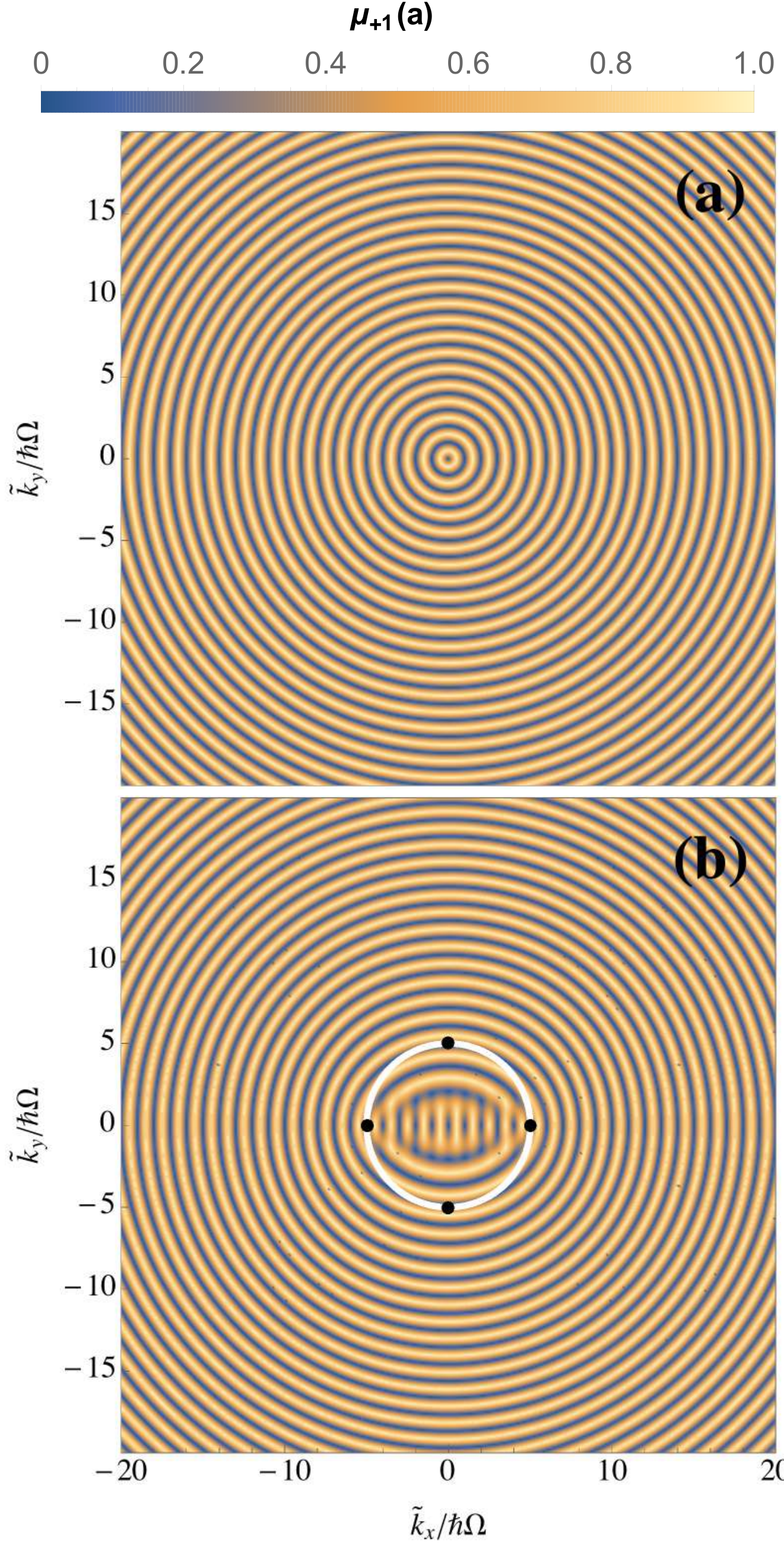} 
\caption{\label{Fig:CharacteristicExponent2D}
Density plot of the characteristic exponent
spectrum $\mu_{+1}(a)$
as a function of $\tilde{k}_x/\hbar\Omega$
and $\tilde{k}_y/\hbar\Omega$ for
(a) $\zeta_x/\hbar\Omega= 10^{-3}$ with
$E_x=10^{-4}\,$V/m and
(b) $\zeta_x/\hbar\Omega=5$ with $E_x=9.7\,$V/m. In 
both cases the frequency of the electromagnetic field has a value $\Omega=50\times10^{9}$ Hz.
The white circle indicates the theoretical
threshold given by the radius $\epsilon=\zeta_x$ at which
there is a transition
from flied-driven strong to weak anisotropy solution.
The black dots denote the limits of the white circle in the 
cases where $\tilde{k}_x/\hbar\Omega=0$ and 
$\tilde{k}_y/\hbar\Omega=0$.}
\end{figure}
\begin{figure}[t]
\includegraphics[width=.48\textwidth]{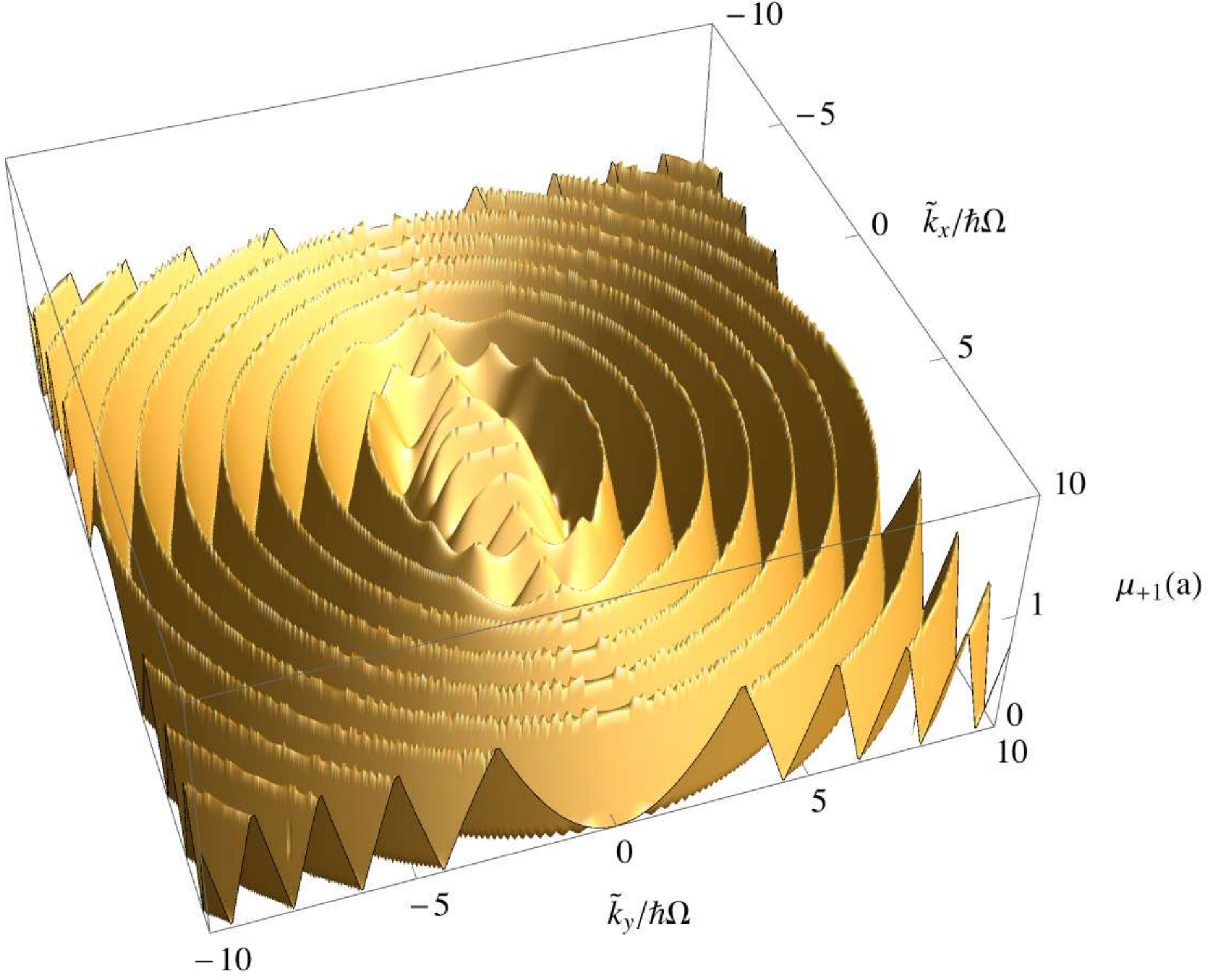}
 \caption{
 Characteristic exponent spectrum $\mu_{+1}(a)$ as a function
 of the normalized 
 momenta $\tilde{k}_x/\hbar\Omega$
and $\tilde{k}_y/\hbar\Omega$.
This plot was made for $\zeta_x/\hbar\Omega=5$ 
 with $E_x=9.7\,$V/m and  $\Omega=50\times10^{9}$ Hz.}
 \label{Fig:CharacteristicExponent3D} 
\end{figure}

\section{Quasienergy spectrum and Wave 
function}\label{sec:QuasienergySpectrumWaveFunction}

\subsection{Quasienergy spectrum}

In this section, we analyse the quasienergy spectrum
produced by the Hill equation \eqref{ec:HillCompoentEta},
and its relations with the stability of its solutions.
The determination of the stability regions of this differential equation 
is quite challenging mainly due to the
imaginary coefficient $q_3$.
The real coefficients $a$, $q_1$ and $q_2$ alone give rise
to the Whittaker-Hill equation \cite{urwin1970iii}, widely discussed
in the spectral theory of periodic differential equations.
The imaginary coefficient $q_3$, however,
introduces additional difficulties that
are rarely addressed in the literature
\cite{ZIENER20124513}.
Among other things, it yields complex characteristic values.
Despite the added complexity, Eq. \eqref{ec:HillCompoentEta}
may be approached by Whitakker's original assumption
\cite{magnus2013hill,Lachapelle_2009} that the solution
should take the Floquet normal form
\begin{equation}\label{ec:hillplus}
\chi_{\eta}(\phi)=\exp\left[i \mu_\eta \phi\right] u_{\eta}(\phi),
\end{equation}
due to the periodicity of the Hamiltonian. Also,
according to the Floquet theorem,
$u_{\eta}(\phi)$ must be a function with period $\pi$.
The function
$\mu_{\eta}(a)$ is termed the characteristic exponent.
As detailed in the following sections, the
characteristic exponent and the quasienergy
spectrum are closely connected.
Thanks to the periodicity of the $u_\eta(\phi)$ function,
$\chi_{\mu}(\phi)$ might be expressed as
a Fourier series expansion. In this manner,
the second-order differential
equation with time-variable periodic coefficients 
is traded for
a matrix eigenvalue problem. The eigenvalues
that stem from it are
precisely the characteristic exponents $\mu_{\eta}(a)$.
These have the form
\begin{equation}\label{ec:CharacteristicExponent}
\mu_{\eta}(a)
=\frac{1}{\pi} \cos^{-1} 
\left[1+\Delta_{\eta}(0)
\left( \cos(\sqrt{a}\pi) -1 \right)\right],
\end{equation}
where $a$ and $\Delta_\eta(0)$  are given by Eqs. 
\eqref{ec:Hill_a} and \eqref{ec:DeterminantDelta},
respectively. For a detailed calculation of
$\mu_{\eta}(a)$ refer to Appendix \ref{AppendixA}.

In general, the solutions of the Hill equation 
\eqref{ec:HillCompoentEta}
fall either on stable ($\mathrm{Im}[\mu_{\eta}(a)] =0$)
or unstable regions ($\mathrm{Im}[\mu_{\eta}(a)] \ne 0$)
depending on the values taken by the coefficients $a$, $q_1$,
$q_2$ and $q_3$
\cite{urwin1970iii,ZIENER20124513,Lachapelle_2009}. 
Surprisingly, the solutions mostly
fall on the stable regions when these coefficients are
restricted by the parametrization of
Eqs. \eqref{ec:Hill_a}-\eqref{ec:Hill_q3}
for seemingly arbitrary domain spaces of
$\tilde{k}_x$, $\tilde{k}_y$ and $\zeta_x$.
Fig. \ref{Fig:CharacteristicExponent2D}
shows a density plot the characteristic exponent 
spectrum for $\mu_{+1}(a)$ as a function of
the normalized momenta $\tilde{k}_x/\hbar\Omega$
and $\tilde{k}_y/\hbar\Omega$. Over the entire parameter
space covered in this figure the solutions
of the Hill equation are stable.

One of most striking features of the spectrum
emerges when one compares
the quasienergies at low and high electric field amplitudes.
The contrast between the effects of
weak ($E_x= 10^{-4}\,$V/m, $\zeta_x/\hbar\Omega=10^{-3}$)
and strong  ($E_x=9.7\,$V/m, $\zeta_x/\hbar\Omega=5$)
electric field amplitudes is shown in
Figs. \ref{Fig:CharacteristicExponent2D} (a)
and (b) respectively.
In the weak electric field regime
(see Fig. \ref{Fig:CharacteristicExponent2D} (a))
the isolines of the characteristic exponent $\mu_{+1}(a)$
as a function of the normalized momenta
form a pattern of concentric circles,
In the strong electric field regime
(see Fig. \ref{Fig:CharacteristicExponent2D} (b))
a spectrum of vertical lines, perpendicular
to the electric field direction, emerges
close to the Dirac point ($\tilde{k}_x=\tilde{k}_y=0$).
The vertical grill is surrounded by an elliptical
outline that is approximately delimited by the
white circle of radius $\epsilon=\zeta_x$.
In the vicinity of this contour ($\epsilon>\zeta_x$)
the spectrum takes on
an elliptical form but further out ($\epsilon \gg \zeta_x$)
the spectrum recovers the circular shape
observed in the low electric field regime.

The appearance of the vertical grill in the
quasienergy spectrum is due to the interplay between
the terms $\cos(2\phi)$, $\sin(2\phi)$ and $\cos(4\phi)$
in Eq. \eqref{ec:HillCompoentEta}.
For example, $\epsilon<\zeta_x$ implies that
$q_2>\left\vert q_3\right\vert$ and $\left\vert q_3\right\vert>q_1$,
therefore the most significant term is $\cos (4\phi)$.
By contrast, $\epsilon>\zeta_x$ implies that
$q_1>\left\vert q_3\right\vert$ and $\left\vert q_3\right\vert>q_1$,
and the dominant element is  $\cos (2\phi)$.  

Further details of the spectrum are appreciated in 
Fig. \ref{Fig:CharacteristicExponent3D} where we show a 3D plot
of the quasienergy $\mu_{+1}(a)$ as a function
of $\tilde{k}_x$ and $\tilde{k}_y$ in the high electric field
regime ($\zeta_x/\hbar\Omega=5$).

\begin{figure}[t]
\includegraphics[width=.5\textwidth]{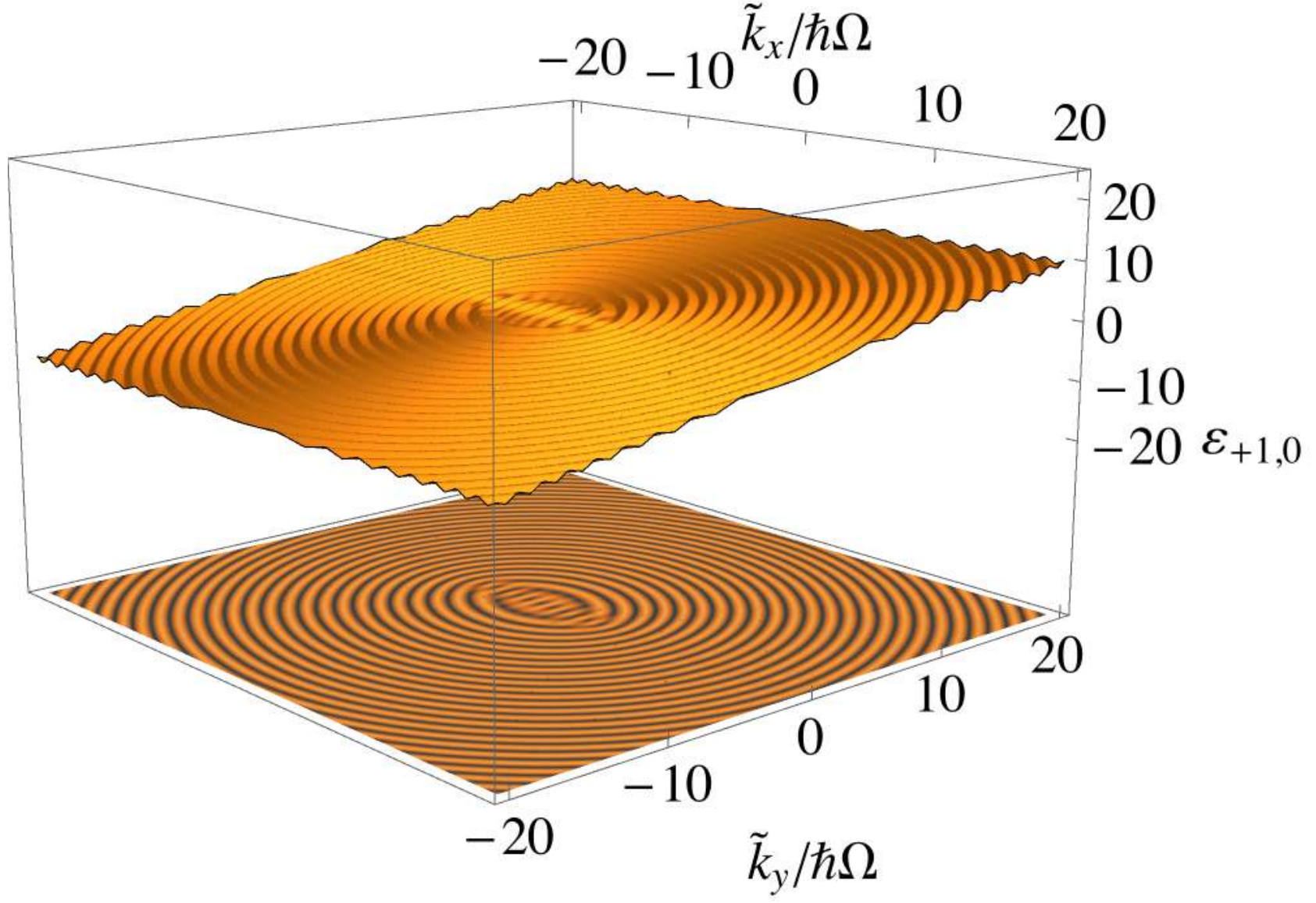}
\caption{Quasienergy $\mathcal{E}_{+1,0}$ spectrum 
as a function of the normalized momenta $\tilde{k}_x/\hbar\Omega$
and $\tilde{k}_y/\hbar\Omega$ in the strong electric field
regime ($\zeta_x/\hbar\Omega=5$, $E_x=9.7\,$V/m)
for $\Omega=50\times10^{9}\,$Hz.
The density plot at the bottom of the plot
is a projection of the 3D plot at the top.
\ref{Fig:CharacteristicExponent2D}(b).}\label{Fig:TiltedQuasiEnergy}
\end{figure}

\subsection{Wave function and Floquet spectrum}
As mentioned above,
the wave function in \eqref{ec:DiracEquationOne} must satisfy the
Floquet theorem as a result of the time periodicity of
the Hamiltonian \eqref{ec:DiracHamiltonianUnderElectromagneticField}.
Hence, following the Floquet theorem, Eqs. \eqref{ec:SolutionOne},
\eqref{ec:suprimdiag}
and \eqref{ec:hillplus} can be combined into the wave function
\begin{equation}
\label{ec:WaveFuncTDC}
\boldsymbol{\Psi}(\phi)
 =\mathcal{N} \exp\left[-\frac{i}{\hbar}\left(\frac{\pi}{4} 
\right)\hat{\sigma}_{y}\right] \mathcal{U}(\phi)\bm{\chi}(0),
\end{equation}
where $\mathcal{N}$ is a normalization constant, 
$\bm{\chi}(0)=(\chi_{+1}(0),\chi_{-1}(0))^{\top}$ is the
initial state vector in
\eqref{ec:HillCompoentEta} and $\mathcal{U}(\phi)$ denotes the time
evolution operator
such that $\chi(\phi)=\mathcal{U}(\phi)\chi(0)$.
As we pointed out before, because $\mathbbm{F}(\phi)$ is a diagonal matrix,
the evolution operator must also be
diagonal. Thus, it can be expressed
quite generally as
\begin{equation}
\mathcal{U}(\phi)=
\begin{pmatrix}
\exp[-i \varepsilon_{+1} \phi]u_{+1}(\phi) & 0 \\
0& \exp[-i \varepsilon_{-1} \phi]u_{-1}(\phi)
\end{pmatrix},
\end{equation}
where $u_{-1}(\phi)=u_{+1}^*(\phi)$ have period $\pi$ and
\begin{equation}
\varepsilon_{\eta}= 2\frac{v_t}{v_y}\frac{ \tilde{k}_y}{\hbar \Omega}
+ \mu_\eta(a).
\end{equation}
It is easy to verify that
the wave function (\ref{ec:WaveFuncTDC}) reduces to the
free-particle wave function (\ref{ec:WaveFunctionFree})
when the electric field vanishes.
From the Floquet theory
\cite{Zhou2011,bukov2015universal,gritsev2017integrable,sandoval2019method},
the time evolution operator is periodic
$\mathcal{U}(\phi)=\mathcal{U}(\phi+l\pi)$
and quasienergy can be expressed as
\begin{eqnarray}\label{ec:QuasiEnergyComplete}
\mathcal{E}_{\eta,l}=\varepsilon_{\eta}+l
 =2\frac{v_t}{v_y}\frac{ \tilde{k}_y}{\hbar \Omega}+ \mu_{\eta,l}(a),
\end{eqnarray}
where
\begin{equation}\label{ec:CharacteristicCoeficientBrillion}
 \mu_{\eta,l}(a)=\mu_\eta(a)+l, \,\,\, l\in \mathbbm{Z},
\end{equation}
is the characteristic exponent
for different Brillouin zones which are tagged by
the integer subscript $l$.

In Fig. \ref{Fig:TiltedQuasiEnergy}, we show the quasienergy for 
$\mathcal{E}_{+1,0}$.
At the bottom of this figure, a density plot of $\mu_{+1,0}(a)$
is shown for reference. One notes that the spectrum consists of
the quasienergies $\mu_{+1}(a)$, plotted in Fig.
\ref{Fig:CharacteristicExponent3D}, on to the tilt
that comes from the first term of Eq. \eqref{ec:QuasiEnergyComplete}.
It arises from the anisotropic character of the
Hamiltonian \eqref{ec:AnsotropicDiracHamiltonian}.
Even though it strongly distorts the symmetry
of the Dirac cone, it has been shown that
interband transitions are not affected by it
in the zero-temperature limit \cite{HerreraNaumisKubo2019}.

\begin{figure}[t]
\includegraphics[width=.2385\textwidth]{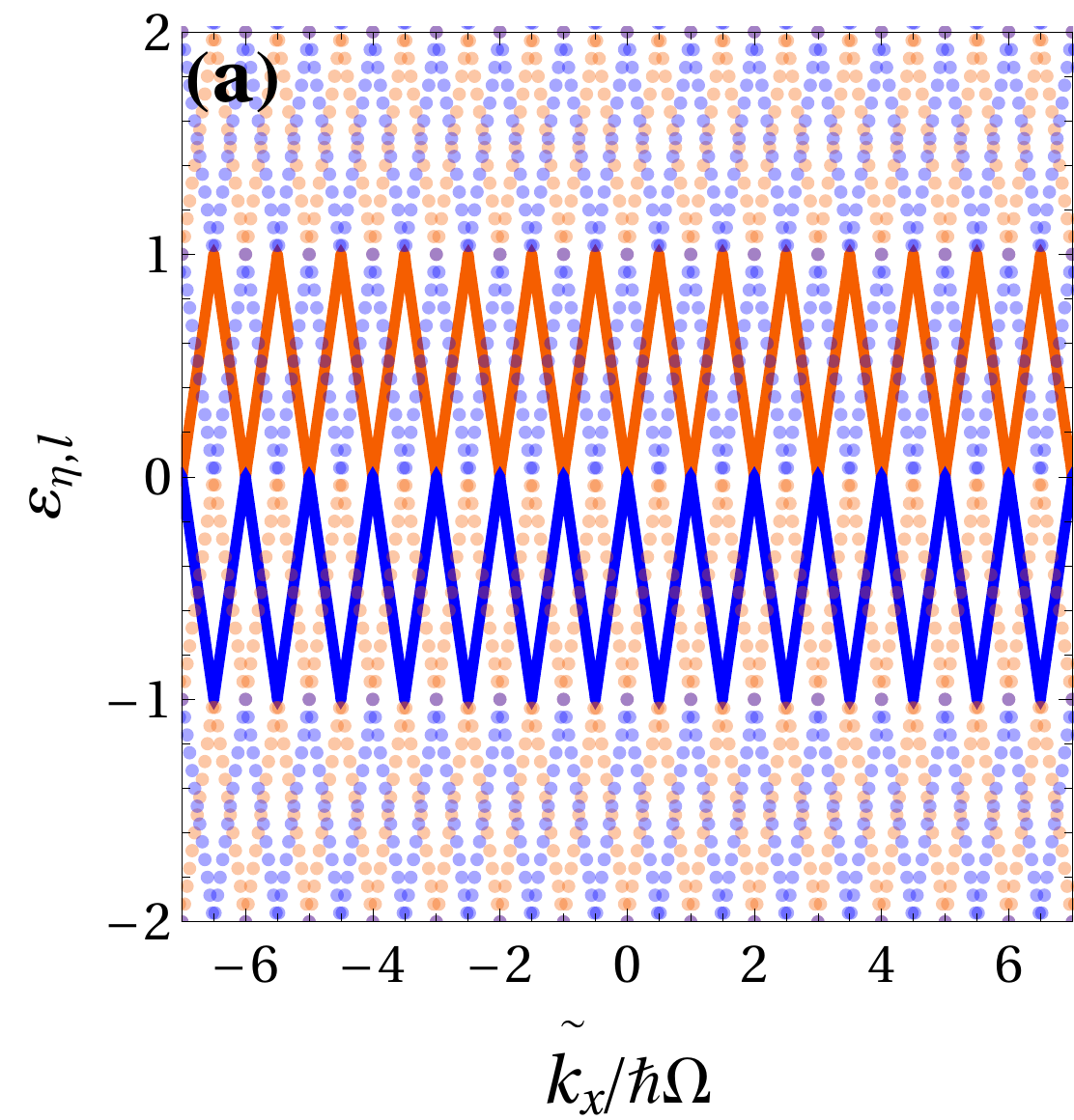}
\includegraphics[width=.2385\textwidth]{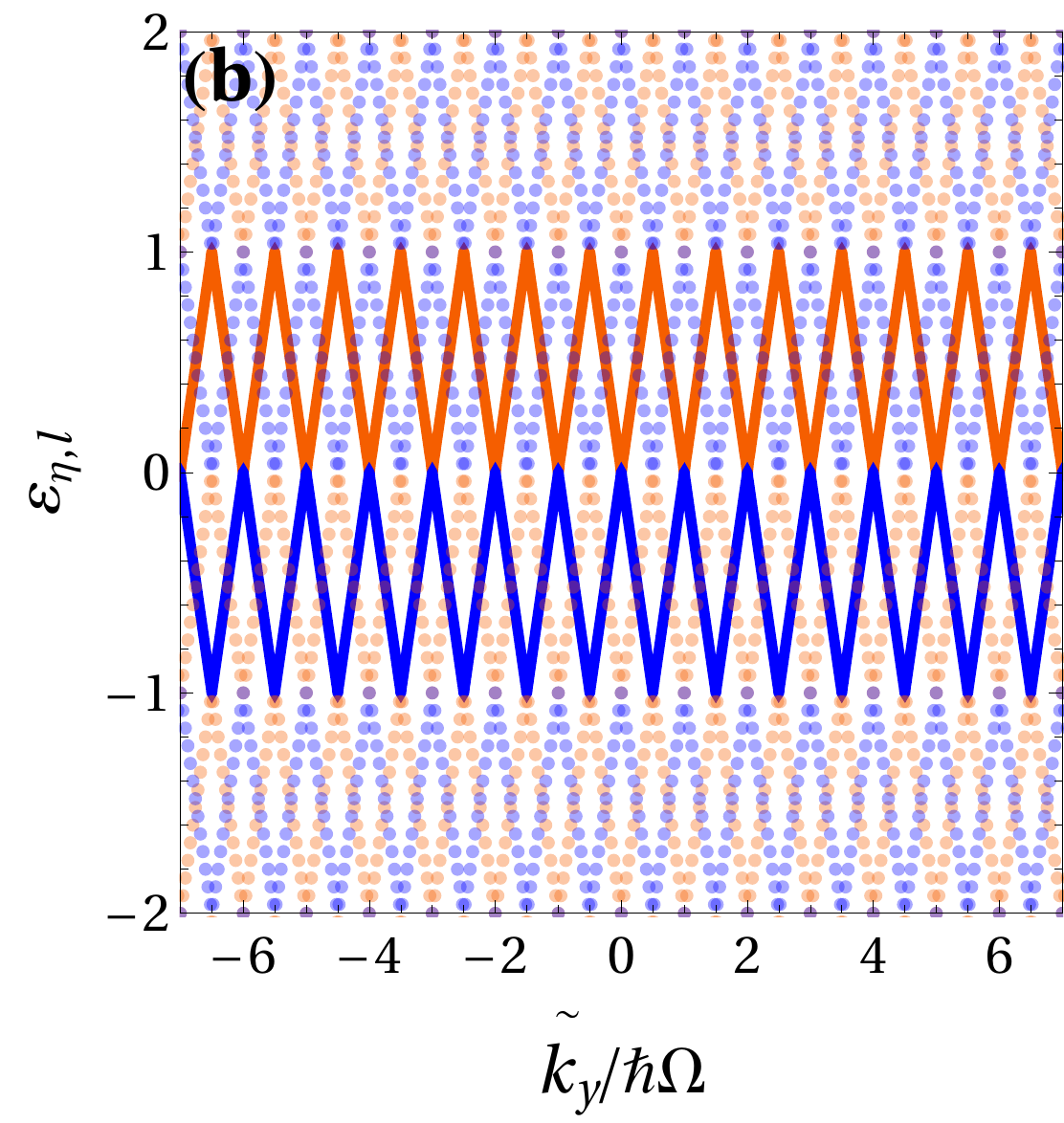}
\includegraphics[width=.2385\textwidth]{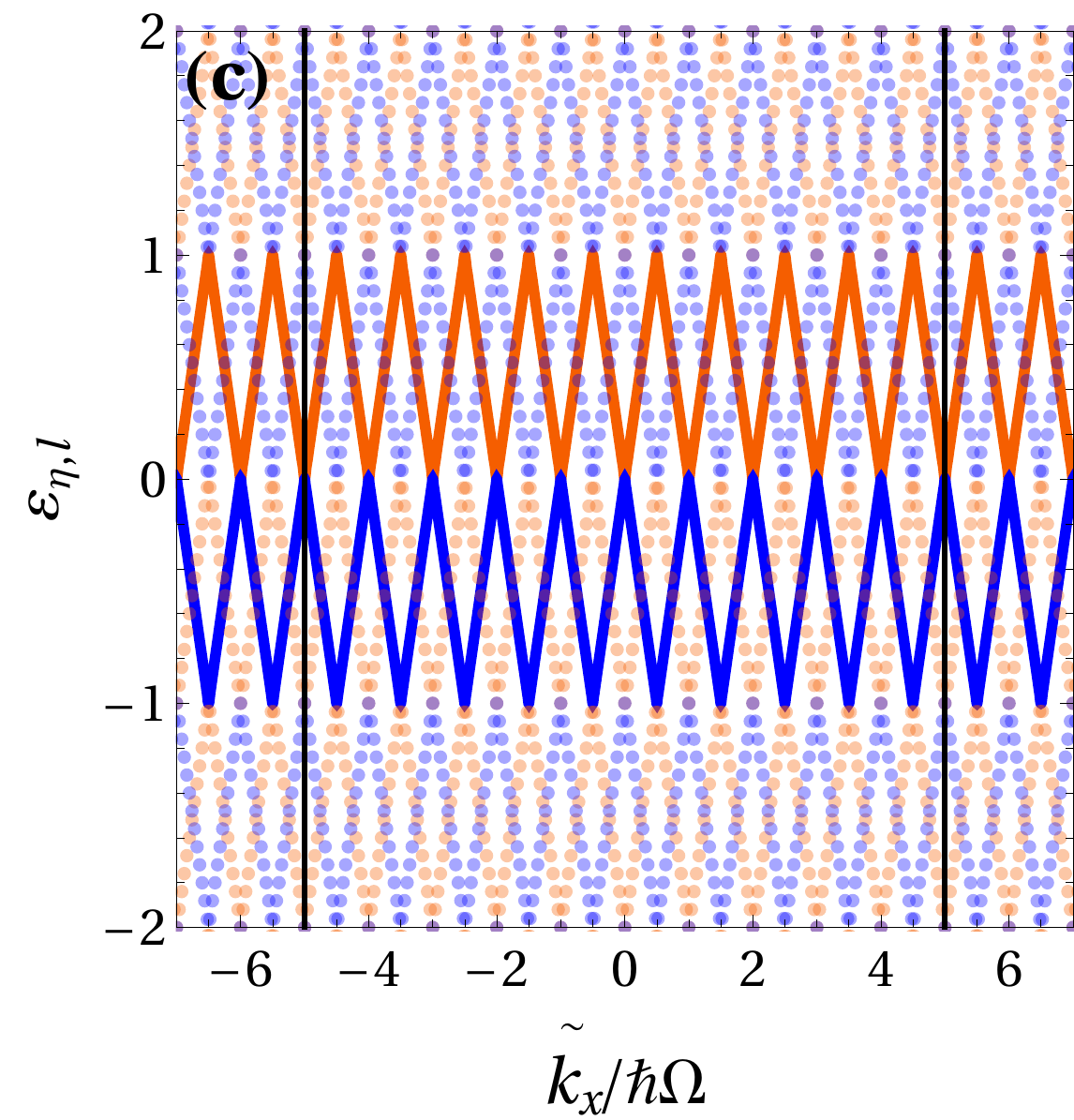}
\includegraphics[width=.2385\textwidth]{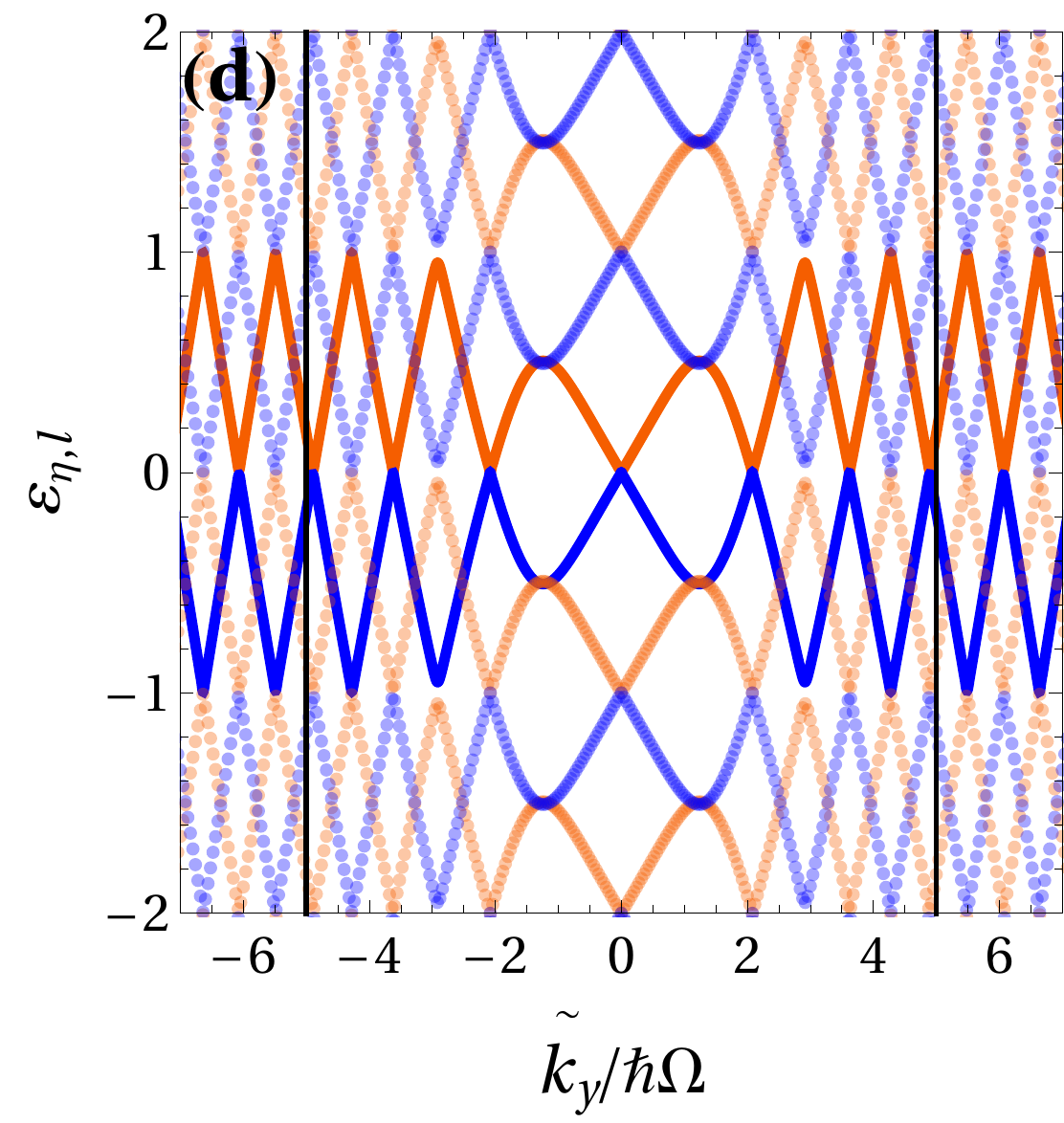}
\caption{
Section planes of the quasienergy $\mu_{\eta,l}(a)$
as a function of $\tilde{k}_x/\hbar \Omega$ or $\tilde{k}_y/\hbar \Omega$
for the Brillouin zones $l=-1,0,1$.
Panels (a) and (b) correspond to the weak electric field
regime ( $\zeta_x/\hbar\Omega= 10^{-3}$, $E_x=10^{-3}\,$V/m,
and $\Omega=50\times10^{9}\,$Hz). The strong electric field
regime ($\zeta_x/\hbar\Omega=5$, $E_x=9.7\,$V/m
and $\Omega=50\times10^{9}\,$Hz) is shown in panels (c) and (d).
The solid orange and blue lines correspond to the quasienergies
$\mu_{+1,0}(a)$ and $\mu_{-1,0}(a)$, respectively.
The light dots plot the quasienergy $\mu_{\pm 1,\pm 1}(a)$
in the adjacent Brillouin zones ($l=\pm 1$).
The threshold $\epsilon=\zeta_x$, where the energy of
the electric field $\zeta_x$ is identical to the
unperturbed energy $\epsilon$,
is indicated by the vertical black lines in panels (c) and (d).
}\label{Fig:CutCuasiEnergy}
\end{figure}

Different cross sections of the quasienergy spectrum $\mu_{\eta,l}(a)$ 
are shown in Fig. \ref{Fig:CutCuasiEnergy}
for the Brillouin zones $l=-1,0,1$.
Panels (a) and (b) show the spectrum section planes
$\tilde{k}_y=0$ and $\tilde{k}_x=0$, respectively,
in the  weak electric field regime
( $\zeta_x/\hbar\Omega= 10^{-3}$, $E_x=10^{-3}\,$V/m,
and $\Omega=50\times10^{9}\,$Hz) .
In this case, both $\tilde{k}_y=0$ and $\tilde{k}_x=0$
cross sections are almost identical since
the electric field is not intense enough to provoke
any distortion to the free particle spectrum.
The quasienergies for the many Brillouin zones
as functions of $\tilde{k}_x$ [panel (a)]
or $\tilde{k}_y$ [panel (b)] have the form of a triangular function.
In the strong electric field regime
($\zeta_x/\hbar\Omega=5$, $E_x=9.7\,$V/m
and $\Omega=50\times10^{9}\,$Hz)
both cuts are radically different.
In panel (c) we note that the $\tilde{k}_y=0$ cross section
of the spectrum resembles those corresponding
to the weak electric field regime also exhibiting
a triangular shaped function of $\tilde{k}_x$.
However, in the $\tilde{k}_x=0$ plane [panel (d)]
the presence of the electromagnetic field
becomes evident as the spectrum is warped
approximately in the domain $-2 < \tilde{k}_y < 2$.
At both ends of this range the quasienergy abruptly
recovers the triangular feature that characterizes
the spectrum in the absence of electromagnetic radiation.
This distortion is roughly bounded by the two black
vertical lines that correspond to the black
dots in Fig. \ref{Fig:CharacteristicExponent2D} (b),
where the strength of the electromagnetic field is comparable
to the energy of the unperturbed system, i.e. $\epsilon=\zeta_x$.

A quite robust feature of the spectrum is
the preservation of the gapless Dirac cone
in the vicinity of the Dirac point:
the quasienergies $\mu_{+1,l}$ and $\mu_{-1,l}$ as
well as $\mathcal{E}_{+1,l}$ and $\mathcal{E}_{-1,l}$ touch
at the tip of the Dirac point despite the 
intensity of the linearly polarized electromagnetic field.
Nevertheless, the conic dispersion relation is stretched
along the $k_y$ direction as a result of the
renormalization of the $v_y$ component of the velocity
due to electronic dressing.
Despite the absolute absence of a gap
a the tip of the Dirac cone,
it is possible  to open up gaps in other
zones of the spectrum.
In Fig. \ref{Fig:CutCuasiEnergy} (d),
the appearance of a small gap
between
$\mu_{+1,l}$ and $\mu_{-1,l}$ can be appreciated 
at $\tilde{k}_y/\hbar \Omega\approx \pm 2$.

\section{Conclusions}\label{sec:Conclusions}

We investigated the quasienergy spectrum of
an anisotropic tilted Dirac material subject to an arbitrarily
intense linearly polarized electromagnetic field.
To this end,  we studied the behavior of 8-$Pmmn$ borophene under
the normal incidence of a linearly polarized field.
We worked out the time-dependent wave function
and the quasienergy spectrum from the Sch\"oendiger equation
via the Floquet theory.
The quasienergy spectrum exhibits a sharp difference between the
weak and strong electromagnetic field regimes. While in the first
the quasienergy as a function of the quasimomenta
($\tilde{k}_x$ and $\tilde{k}_y$)
is highly isotropic the latter presents an anisotropic
pattern in the low energy region that resembles a grid
aligned perpendicularly to the direction of the
radiation's electric field.
This pattern abruptly disappears beyond the
threshold where the free kinetic energy
of the carriers is larger than the energy associated
to the electric field.
Near this threshold, a gap opens up.
Probably the most astonishing feature of this
spectrum is that, eventhough the gapless
Dirac cone is preserved, the dispersion relation
is stretched along the direction perpendicular
to the field's polarization. This is an outcome
of the electronic dressing and the consequent 
rescaling of the velocity.
This mechanism could be exploited to tune
the electronic properties of Dirac materials
through the field parameters.
A fundamental aspect, yet to be addressed, is
the capability of circularly polarized fields
to adjust these properties.

\section{Author's Contributions}

All authors contributed equally to this work.

\section{Acknowledgements}\label{sec:Acknowledgements}

This work was supported by DCB UAM-A grant numbers
2232214 and 2232215, and UNAM DGAPA PAPIIT IN102620.
 J.C.S.S. and V.G.I.S acknowledge the total support from 
DGAPA-UNAM fellowship. 

\section{Data Availability}

The data that support the findings of this study are available from the corresponding author
upon reasonable request

\appendix

\section{}\label{AppendixA}

This Appendix covers the method developed by  Whittaker
\cite{magnus2013hill,Lachapelle_2009}
to determine the characteristic exponent
\eqref{ec:CharacteristicExponent}.
Eq. \eqref{ec:HillCompoentEta} is the starting point.
This equation is periodic
and therefore its solution must comply with the
Floquet theorem. Thus, the solution is given by
\begin{equation}
\chi_{\eta}(\phi)=e^{i\mu_{\eta}\phi}u_{\eta}(\phi) \, ,
\end{equation}
where $\eta=\pm 1$, $u_{\eta}(\phi)$ is a function with period $\pi$
and $\mu_\eta$ denotes the characteristic exponent.
Thanks to the periodicity of $u_{\eta}(\phi)$, the
wave function can
be expanded as a Fourier series as
\begin{equation}
\chi_{\eta}(\phi)=\mathrm{e}^{i\mu_{\eta} \phi} \sum_{r=-\infty}^{\infty}C^{(\eta)}_{2r}
\mathrm{e}^{i 2r \phi} \, .
\end{equation}
Substituting the preceding equation into \eqref{ec:HillCompoentEta}
and rearranging  the coefficients,
we obtain the following recurrence relation
\begin{multline}\label{ec:RecurenceRelation}
\gamma_{2r}C^{(\eta)}_{2(r-2)}+\alpha_{2r}C^{(\eta)}_{2(r-1)}+C^{(\eta)}_{2r}\\
+\beta_{2r}C^{(\eta)}_{2(r+1)}+\gamma_{2r}C^{(\eta)}_{2(r+2)}=0,    
\end{multline}
where 
\begin{equation}
\alpha_{2r}=\frac{1}{2}\frac{q_1-i\eta q_3 }{a-(\mu+2r)^2},
\end{equation}
\begin{equation}
\beta_{2r}=\frac{1}{2}\frac{q_1+i\eta q_3 }{a-(\mu+2r)^2},
\end{equation}
\begin{equation}
\gamma_{2r}=\frac{1}{2}\frac{q_2}{a-(\mu+2r)^2}.
\end{equation}
The equation parameters $a$, $q_1$, $q_2$ and $q_3$ are defined in  Eqs.
\eqref{ec:Hill_a}-\eqref{ec:Hill_q3}.
The recurrence relation \eqref{ec:RecurenceRelation} can be put in the form
of a linear equation as
\begin{equation}
\mathcal{A}_{r}\left(\mu,\eta,a,q_1,q_2,q_3\right)\boldsymbol{C}^{(\eta)}=0
\end{equation}
where
$\boldsymbol{C}^{(\eta)}
=\left(C^{(\eta)}_2,C^{(\eta)}_4,C^{(\eta)}_6,...\right)^{\top}$
and
\begin{eqnarray}
&&\mathcal{A}_{r}\left(0,\eta,a,q_1,q_2,q_3\right)=\nonumber\\
&&
\begin{pmatrix}
1 & \alpha_{2r} & \gamma_{2r}& 0&0&0&0&0&0\\
\beta_{2r-2}&. & . & . & .& .&.&.&0\\
0&. & 1 & \alpha_{4} & \gamma_{4}& 0&0&.&0\\
0&. & \beta_{2} & 1 &  \alpha_{2} & \gamma_{2}&0&.&0\\
0&. & \gamma_{0} & \beta_{0} &  1& \alpha_{0} & \gamma_{0}&.&0\\
0&. & 0 & \gamma_{-2} & \beta_{-2}& 1&\alpha_{-2} & .&0\\
0&. & 0 & 0&\gamma_{-4} & \beta_{-4}&1&.&0\\
0&. & . & . &. & . & .&.&\alpha_{-2r+2}\\
0&0&0&0 & 0&0&\gamma_{-2r} & \beta_{-2r}&1
\end{pmatrix}.\nonumber\\
\end{eqnarray}
To avoid the trivial solution we demand that the determinant of
the precedent matrix vanishes:
\begin{equation}\label{ec:DeterminantDelta}
\Delta_{\eta}(0)
=\mathrm{det}\left[\mathcal{A}_{r}\left(0,\eta,a,q_1,q_2,q_3\right)\right]
=0.
\end{equation}
It can be proven that this determinant may be written in the
compact form\cite{Lachapelle_2009}
\begin{equation}
\sin^2\left(\mu_\eta(a) \frac{\pi}{2}\right)
=\Delta_{\eta}(0)\sin^{2}\left(\sqrt{a}\frac{\pi}{2}\right)\, .
\end{equation}
Solving the above equation for $\mu_\eta(a)$
we obtain
\begin{equation}
\mu_\eta(a)= \frac{1}{\pi} \cos^{-1}
\left[1+\Delta_{\eta}(0)(\cos(\sqrt{a}\pi)-1) \right] .
\end{equation}
This expression presents the advantage
of being efficiently evaluated by
using numerical methods. Strictly speaking
$\mathcal{A}_{r}$
is infinite-dimensional, however good numerical
convergence of $\mu_\eta(a)$ is achieved by
cutting it down to a $400\times 400$ matrix.




\bibliography{biblio.bib}

\end{document}